\documentclass[aps,prl,reprint,superscriptaddress,hyperlinks]{revtex4-2}
\usepackage{microtype}
\usepackage{chemformula} 
\usepackage[T1]{fontenc} 
\usepackage{physics}
\usepackage{subcaption}
\usepackage{float}
\usepackage[colorlinks=true,linkcolor=blue,anchorcolor=blue,citecolor=blue,urlcolor=blue,bookmarks=true]{hyperref}
\usepackage[capitalise]{cleveref}

\newcommand{\ms}{\mathbf{S}}

\begin{document}

\title{Reversible nonrelativistic magnon spin transport in ferroelastic altermagnets}

\author{Haozhou Cai}
\affiliation{State Key Laboratory of Low Dimensional Quantum Physics and Department of Physics, Tsinghua University, Beijing 100084, China}

\author{Jian Wu}
\affiliation{State Key Laboratory of Low Dimensional Quantum Physics and Department of Physics, Tsinghua University, Beijing 100084, China}
\affiliation{Frontier Science Center for Quantum Information, Beijing, China}

\author{Weiyi Pan}
\email{Weiyi.Pan@physik.uni-regensburg.de}
\affiliation{State Key Laboratory of Low Dimensional Quantum Physics and Department of Physics, Tsinghua University, Beijing 100084, China}
\affiliation{Institute for Theoretical Physics, University of Regensburg, 93040 Regensburg, Germany}

\date{\today}

\begin{abstract}
	Magnons in antiferromagnetic (AFM) insulators facilitate low-dissipation, stray-field-free spin transport. However, achieving nonvolatile, field-free control over magnon spin currents remains elusive. Here, based on symmetry analysis, we propose a universal mechanism for the active manipulation of magnon spin transport via ferroelastic transitions in two-dimensional (2D) altermagnets (AMs)---a class of unconventional AFMs simultaneously exhibiting compensated magnetization and nonrelativistic spin splitting. We show that these transitions effectively reorient principal crystal axes and modulate the underlying magnetic exchange anisotropy. Consequently, this magnetoelastic coupling drives nonrelativistic anisotropic spin transport that is ferroelastically switchable without the need for external magnetic fields or Berry curvature, leading to sign reversals in the spin Seebeck and spin Nernst conductivities. We validate this mechanism using first-principles calculations and spin-model analyses of an AM \ch{CoTe2} monolayer. Our findings establish a symmetry-based magnetoelastic paradigm for the nonvolatile control of magnon spin transport in 2D AMs, opening new avenues toward energy-efficient, reconfigurable AFM magnonic devices.

\end{abstract}

\maketitle

\textit{Introduction}---Magnons, the quanta of spin-wave excitations, can carry spin angular momentum in insulating systems without charge motion, offering a low-dissipation, high-frequency alternative to charge-based electronics \cite{chumakMagnonTransistorAllmagnon2014,chumakMagnonSpintronics2015,flebus2024MagnonicsRoadmap2024}. A promising platform for magnonics is the collinear antiferromagnet (AFM), whose zero net magnetization ensures immunity to stray fields and facilitates high-density integration. Moreover, its chiral magnon modes ($\alpha$ and $\beta$), which carry opposite spin angular momenta ($\mp\hbar$), enable chirality-based spintronics \cite{chengSpinPumpingSpinTransfer2014,baltzAntiferromagneticSpintronics2018,rezendeIntroductionAntiferromagneticMagnons2019,flebusMagnonicsCollinearMagnetic2021,wangMagnonsVanWaals2026}. 
However, in conventional AFMs, the degeneracy of chiral magnon modes causes their spin transport contributions to cancel, and therefore suppresses Joule-heating-free spin transport phenomena mediated by charge-neutral magnons, such as the spin Seebeck effect (SSE) and the spin Nernst effect (SNE). Conventional approaches to realizing nontrivial spin transport in AFM systems rely either on lifting the degeneracy of chiral magnon modes with external magnetic fields \cite{sekiThermalGenerationSpin2015,rezendeTheorySpinSeebeck2016,liSpinSeebeckEffect2019}, or on exploiting the intrinsic SNE arising from magnon Berry curvature induced by spin-orbit coupling (SOC) \cite{chengSpinNernstEffect2016,zyuzinMagnonSpinNernst2016,leeMagnonicQuantumSpin2018,bazazzadehSymmetryEnhancedSpinNernst2021}. These field- and SOC-dependent strategies complicate device design and constrain the pool of candidate materials. 
Recently, altermagnets (AMs) \cite{yuanGiantMomentumdependentSpin2020,maMultifunctionalAntiferromagneticMaterials2021,mazinEditorialAltermagnetismANew2022,smejkalConventionalFerromagnetismAntiferromagnetism2022,songAltermagnetsNewClass2025,huangSpinInversionEnforced2025}, whose macroscopic magnetic compensation is protected by crystal rotation symmetry rather than the $\mathcal{P}\mathcal{T}$ or $\tau U$ (translation combined with spin rotation) symmetry typical of conventional AFMs, have bypassed the aforementioned constraints via nonrelativistic chiral magnon splitting \cite{smejkalChiralMagnonsAltermagnetic2023,liuChiralSplitMagnon2024,hoyerAltermagneticSplittingMagnons2025,chenUnconventionalMagnonsCollinear2025a,sunObservationChiralMagnon2025,xieGeneralTheoryChiral2026}, thereby enabling field-free magnon spin transport \cite{cuiEfficientSpinSeebeck2023,yuChiralMagnonsAltermagnetic2024,weissenhoferAtomisticSpinDynamics2024,yangAltermagnetdrivenMagnonSpin2026}.

However, generating field-free magnon spin currents is insufficient for practical magnonic devices. For functional applications, such as nonvolatile memories and magnonic logic devices based on magnon-induced spin torque \cite{chumakMagnonSpintronics2015,wangMagnetizationSwitchingMagnonmediated2019,huangManipulatingChiralSpin2024,zhangElectricalManipulationSpin2025}, the sign of magnon spin current must be switchable in a reversible and nonvolatile manner. Although recent efforts have modulated magnon spin transport via electric fields \cite{liuElectricFieldControl2021,chengNonvolatileMagnonField2024,huangManipulatingChiralSpin2024} and mechanical strain \cite{esterasMagnonStraintronics2D2022,zhouPiezoelectricStrainControlledMagnon2022,yuChiralMagnonsAltermagnetic2024}, these methods lack nonvolatility or fail to achieve sign reversal of spin currents. 
Alternatively, while thermodynamic sublattice compensation and electrostatic strain-mediated switching \cite{gepragsOriginSpinSeebeck2016a,parsonnetNonvolatileElectricField2022,liElectricFieldSwitching2024,wangElectricalExcitationDetection2024,kawamotoUnderstandingSpinCurrents2024} can accomplish this reversal, they are fundamentally limited by the narrow temperature windows near the magnetic compensation point, the fabrication complexity of integrated piezoelectric heterostructures, or the reliance on external bias fields.
Therefore, establishing a robust paradigm for the nonvolatile, field-free control of magnon spin transport remains highly desirable.

\begin{figure*}
	\centering
	\includegraphics[width=\linewidth]{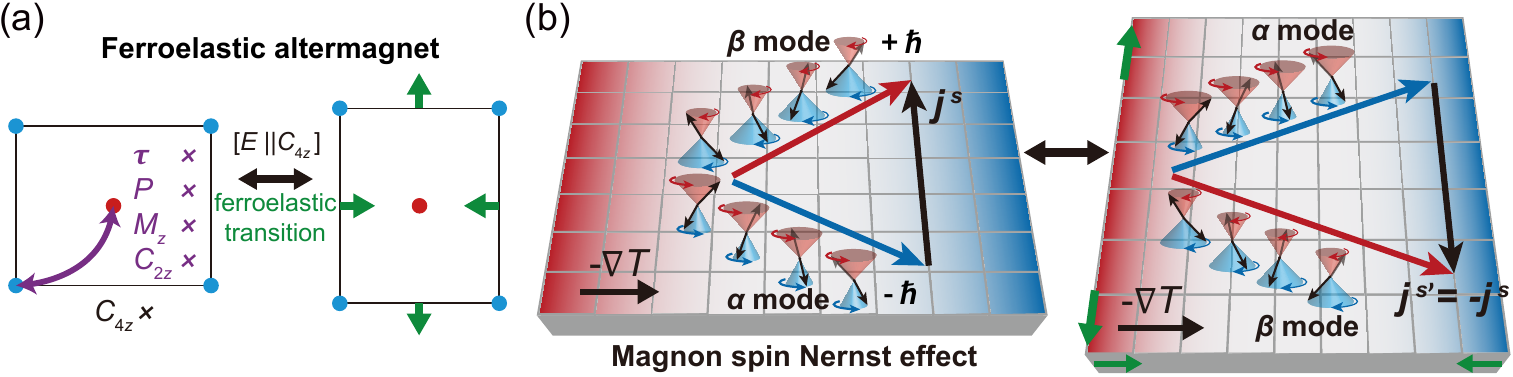}
	\caption{(a) Ferroelastic switching of a rectangular lattice altermagnet, with forbidden symmetries labeled. (b) Ferroelastic switching of the spin Nernst effect under a thermal gradient applied along the $x$-axis. Each cone represents the precession of antiferromagnetically coupled spins on sublattices A and B. The $\alpha$ ($\beta$) mode exhibits collective right (left)-handed precession, carrying $-\hbar~(+\hbar)$ spin angular momentum. Blue (red) arrows indicate the propagation of $\alpha$ ($\beta$) magnons. $\mathbf{j}^s$ and $\mathbf{j}^{s\prime}$ denote magnon spin currents before and after switching, respectively.} \label{fig:concept}
\end{figure*}

Here, we propose a universal, symmetry-guided mechanism for the nonvolatile, field-free control of magnon spin transport through the interplay between ferroelasticity and altermagnetism. Focusing on 2D ferroelastic AMs, we demonstrate that the strain-induced lattice reorientations modifies the anisotropy of magnetic exchange interactions and reshapes the chiral magnon dispersions. This intrinsic magnetoelastic coupling profoundly modulates nonrelativistic magnon spin transport, manifesting as a complete reversal of the anisotropic SSE and SNE.
We explicitly validate this mechanism using first-principles calculations and spin-model analyses of an AM \ch{CoTe2} monolayer. Our findings establish magnetoelastic coupling as a design principle for nonvolatile magnonic devices, broadly applicable to ferroelastic AMs where strain-induced lattice reorientations govern magnon transport properties.

\textit{Symmetry considerations}---To identify 2D AMs enabling nontrivial magnon-mediated spin transport, we begin with a symmetry analysis.
Based on 80 layer groups (LGs) \cite{fuSymmetryClassification2D2024}, 2D altermagnetism is formally defined by the spin layer group $G_s = [E\|H] + [C_2\|G-H]$, where the notation $[R_s\|R_l]$ denotes a combined operation of $R_s$ in spin space and $R_l$ in real space. Here, $H$ is a halving subgroup of the parent layer group $G$, while $E$ and $C_2$ denote the identity operation and a $180^\circ$ rotation around an axis perpendicular to the spins, respectively. To preserve the AM phase by avoiding trivial mappings between opposite-spin sublattices, the coset $G-H$ must exclude translation $\tau$, spatial inversion $\mathcal{P}$, horizontal mirror reflection $M_z$, and out-of-plane twofold rotation $C_{2z}$. Consequently, as previously established \cite{zengDescriptionTwodimensionalAltermagnetism2024}, 2D AMs are classified by seven spin layer Laue groups, which are listed in Table~S3 of the Supplemental Material (SM) \cite{supp}.
Building upon this classification, we evaluate the associated magnon spin transport tensors in the linear response regime. Driven by an applied thermal gradient $\nabla T$, the diffusive motion of magnons generates a spin current $\mathbf{j}^s$, which is given by $j_i^s=\sigma^s_{ij}(-\partial_j T)$. Neglecting SOC-induced Berry curvature effects, the magnon spin conductivity ($\sigma^s$) tensor can be evaluated using semiclassical Boltzmann transport theory under the relaxation-time approximation \cite{cuiEfficientSpinSeebeck2023,yuChiralMagnonsAltermagnetic2024,hopfnerSignChangesHeat2025} (see Sec.~S5 of SM \cite{supp}). Our symmetry analysis (Table~S3 \cite{supp}) reveals that only four spin layer Laue groups ($^22/^2m_x$, $^2m^2m^1m$, $^24/^1m$, and $^24/^1m^2m^1m$) permit a nonvanishing spin conductivity. These groups correspond to rectangular or square crystal systems (\cref{fig:concept}(a)) and are characterized by $d$-wave chiral magnon splitting.

In 2D AMs with nonzero spin conductivity, nonvolatile and reversible control of spin current can be achieved through ferroelastic switching, which is enabled by toggling the spontaneous ferroelastic strain via external stress. Such ferroelasticity originates from a symmetry-lowering transition from a high-symmetry prototypical phase, further limiting material candidates to rectangular systems with unequal lattice constants ($a\ne b$), which belong to LG 8--48 \cite{aizuPossibleSpeciesFerroelastic1969,pengFerroelasticAltermagnetism2025}. Consequently, the relevant ferroelastic AMs are restricted to only two spin layer Laue groups, $^22/^2m_x$ (LG 8--18) and $^2m^2m^1m$ (LG 19--48). Both groups dictate the same purely off-diagonal spin conductivity tensor:
\begin{equation}
	\sigma^s=\begin{pmatrix}
		0             & \sigma_{xy}^s \\
		\sigma_{xy}^s & 0
	\end{pmatrix}.
\end{equation}
In a typical two-state system where the energetically degenerate ferroelastic phases are linked by a spatial fourfold rotation $[E\|C_{4z}]$, the ferroelastic transition transforms the spin conductivity tensor as:
\begin{equation}
		\sigma^{s\prime}=[E\|C_{4z}]\sigma^s[E\|C_{4z}]^{-1} = 
		\begin{pmatrix}
			0    & -\sigma_{xy}^s \\
			-\sigma_{xy}^s & 0
		\end{pmatrix}.
\end{equation}
This relation demonstrates that ferroelastic switching can deterministically reverse the sign of the magnon spin conductivity ($\sigma_{xy}^s \to -\sigma_{xy}^s$). Such magnetoelastic coupling thus provides a deterministic pathway to control magnon spin transport, enabling the reversible switching of the anisotropic SSE and SNE (\cref{fig:concept}(b)), as elaborated later.

\begin{figure}
	\centering
	\includegraphics[width=\linewidth]{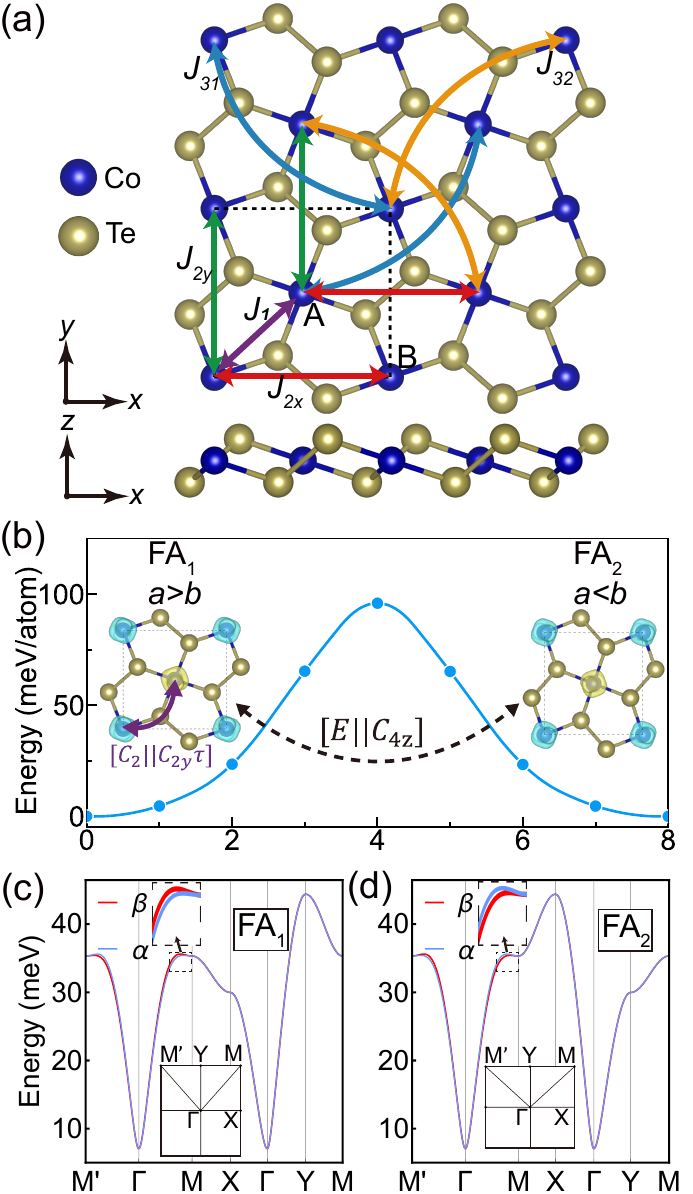}
	\caption{(a) Top and side views of the crystal structures of in \ch{CoTe2}. Double-headed arrows indicate the Heisenberg exchange interaction pairs. (b) Energy barriers along the ferroelastic switching paths. The insets depict the initial (FA$_1$, $a>b$) and final (FA$_2$, $a<b$) states of the switching process, along with symmetry operations relating them. Yellow (blue) isosurfaces denote spin-up (spin-down) densities. (c, d) Chiral magnon bands for the two ferroelastic phases (FA$_1$ and FA$_2$), with insets depicting the first Brillouin zone. Red and blue lines represent the $\beta$ and $\alpha$ modes, respectively.} \label{fig:CoTe2-0}
\end{figure}

\textit{Atomic structures, ferroelastic phases and magnetic properties}---To demonstrate the mechanism of ferroelastically switchable magnon transport in AMs, we use the AM \ch{CoTe2} monolayer as a model system, which crystallizes in a rectangular lattice with LG 17, as depicted in \cref{fig:CoTe2-0}(a). The buckled structure breaks $C_{4z}$ symmetry, resulting in unequal lattice constants $a=6.364$~\AA\ and $b=6.095$~\AA.
The dynamical and mechanical stability of \ch{CoTe2} is confirmed by the absence of imaginary frequencies in its phonon spectrum (Fig.~S2 \cite{supp}) and satisfaction of the Born stability criteria (Table~S1 \cite{supp}).
\ch{CoTe2} hosts two ferroelastic phases (FA$_1$ and FA$_2$) that are mutually switchable via uniaxial strain along the $x$- or $y$-axis. Solid-state nudged elastic band calculations \cite{henkelmanImprovedTangentEstimate2000,sheppardGeneralizedSolidstateNudged2012} reveal a switching barrier of 96~meV/atom (\cref{fig:CoTe2-0}(b)), comparable to those of other ferroelastic materials \cite{dingFerroelasticallyTunableAltermagnets2025,zhangTwoDimensionalFerroelasticSemiconductors2020,wuIntrinsicFerroelasticityMultiferroicity2016}. Geometrically, this ferroelastic transition is equivalent to a $C_{4z}$ operation on the crystal structure, interchanging the principal axes and thereby modulating anisotropic physical properties.

To investigate the magnetic properties of \ch{CoTe2}, we construct a spin model to capture the key magnetic interactions between localized moments on Co ions:
\begin{equation}\label{eq:spinmodel}
	\begin{split}
		H & =\sum_{(ij)_{AB}} J_{ij}^{AB} \ms_{A,i} \cdot \ms_{B,j}+ \sum_{i\in A} K(S_{A,i}^z)^2 + \sum_{i\in B} K(S_{B,i}^z)^2 \\
		  & + \sum_{(ij)_{AA}} J_{ij}^A \ms_{A,i} \cdot \ms_{A,j} + \sum_{(ij)_{BB}} J_{ij}^B \ms_{B,i} \cdot \ms_{B,j}
	\end{split}
\end{equation}
Here, $A$ and $B$ represent the two magnetic sublattices, with $\mathbf{S}_{A(B),i}$ denoting the spin at site $i$ of sublattice $A$ ($B$), normalized to $S=1/2$ for Co ions. The terms involving $J_{ij}^{AB}$ describe the inter-sublattice exchange couplings, whereas those with $J_{ij}^A$ and $J_{ij}^B$ represent the intra-sublattice exchange interactions. The parameter $K$ denotes the single-ion anisotropy energy.

Since the crystal structures of the two ferroelastic phases are related by a $C_{4z}$ operation, we focus our discussion of the magnetic interactions on the FA$_1$ phase. The exchange parameters (denoted by double-headed arrows in \cref{fig:CoTe2-0}(a)) are extracted from first-principles calculations, as detailed in Sec.~S3 of SM \cite{supp}. The magnetism is dominated by the antiferromagnetic nearest-neighbor interaction $J_1=17.69$~meV. The breaking of $C_{4z}$ symmetry induces a pronounced anisotropy in the exchange couplings along the $x$ and $y$ axes ($J_{2x}=3.79$~meV and $J_{2y}=-3.42$~meV). Furthermore, the nontrivial spatial relation between the two sublattices results in inherently inequivalent diagonal intra-sublattice interactions ($J_{31}=-0.22$~meV and $J_{32}=-0.52$~meV). Incorporating an easy-axis magnetic anisotropy ($K=-0.7$~meV), our Monte Carlo simulations confirm a N\'eel AFM ground state below $T_N \approx 34$~K (Fig.~S4 \cite{supp}). In this state, the opposite-spin sublattices in \ch{CoTe2} are related by the spin group operations $ [C_2 \| C_{2y}\tau] $ and $[C_2 \| M_y\tau]$ rather than $[C_2\|\tau]$ or $[C_2\|P]$, classifying the system as an AM \cite{smejkalConventionalFerromagnetismAntiferromagnetism2022}. The electronic bands (Fig.~S1 \cite{supp}) indeed exhibit the hallmark alternating spin splitting along the M$'$-$\Gamma$-M path, which is reversible via the ferroelastic transition. Moreover, the insulating nature of \ch{CoTe2} ensures that its low-energy transport properties are dominated by magnons. In the following, we demonstrate that ferroelasticity can also switch both the chiral magnon splitting and the transport properties.

\begin{figure*}
	\centering
	\includegraphics[width=\linewidth]{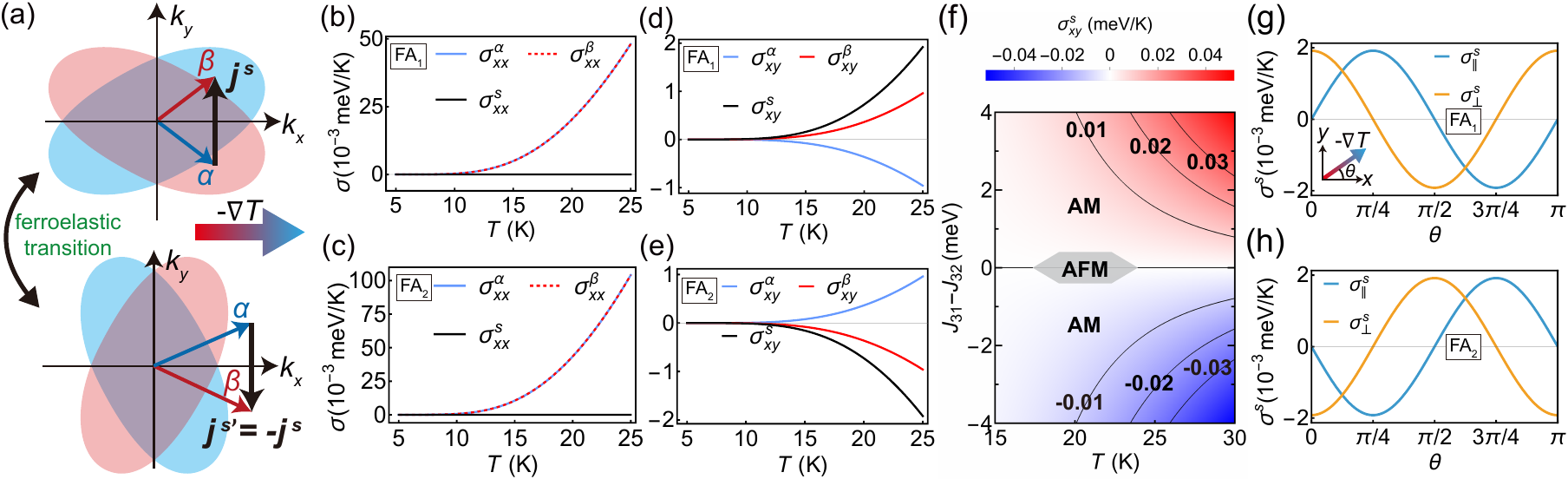}
	\caption{(a) Ferroelastic switching of the spin current under a thermal gradient along the $x$-axis. The net spin current is governed by the imbalance between the $\beta$ and $\alpha$ magnon flows. The blue (red) shaded regions represent energy contours of the $\alpha\,(\beta)$ magnon bands. The blue (red) arrows indicate the flow of $\alpha\,(\beta)$ magnons. The black arrows represent the net spin current $\mathbf{j}^s$. (b, c) Net diagonal spin conductivity $\sigma_{xx}^s$ and the individual contributions from $\alpha$ and $\beta$ modes in the (b) FA$_1$ and (c) FA$_2$ phases. (d, e) Net off-diagonal spin conductivity $\sigma_{xy}^s$ and the chirality-resolved contributions in the (d) FA$_1$ and (e) FA$_2$ phases. (f) $\sigma_{xy}^s$ as a function of temperature and the difference in exchange interactions $J_{31}-J_{32}$. The sum $J_{31}+J_{32}$ is fixed at $-0.74$~meV. (g, h) Anisotropic spin Seebeck ($\sigma_{\parallel}^s$) and Nernst ($\sigma_{\perp}^s$) conductivities at 25~K for the (g) FA$_1$ and (h) FA$_2$ phases of \ch{CoTe2}, respectively. The inset in (g) illustrates the applied thermal gradient $-\nabla T$, which is oriented at angle $\theta$ from the $x$-axis.}\label{fig:CoTe2-1}
\end{figure*}

\textit{Ferroelastically switchable magnon chiral splitting}---Based on the spin model (\cref{eq:spinmodel}), we calculate the magnon band structure within the framework of linear spin-wave theory, as detailed in Sec.~S4 of SM \cite{supp}. The resulting spectrum comprises two chiral magnon modes, $\alpha$ and $\beta$, with dispersions $E_{\alpha,\mathbf{k}}$ and $E_{\beta,\mathbf{k}}$, which carry opposite spin angular momenta of $-\hbar$ and $+\hbar$, respectively \cite{cuiEfficientSpinSeebeck2023}. In the $\text{FA}_1$ phase of $\text{CoTe}_2$, the chiral splitting is given by $\Delta E_k=E_{\beta,k}-E_{\alpha,k}=4S(J_{31}-J_{32})\sin k_x \sin k_y$, indicating that the magnon bands exhibit opposite chiral splitting along the $\Gamma$-M and $\Gamma$-M$'$ directions, as clearly shown in \cref{fig:CoTe2-0}(c) (the dispersion across the entire first Brillouin zone is provided in Fig.~S5 \cite{supp}). The magnitude of the chiral splitting reaches 0.6~meV, comparable to the 0.7~meV predicted for the ferroelectric AM \ch{V2I2O2BrCl} \cite{wangTwodimensionalDualswitchableFerroelectric2025}. 
Furthermore, the $[E\|C_{4z}]$ operation, which effectively governs the transition between the two ferroelastic phases, interchanges the magnon bands along orthogonal directions. This directly induces a reversal of the chiral splitting between the FA$_1$ and FA$_2$ phases (\cref{fig:CoTe2-0}(d)), establishing nonvolatile ferroelastic control over magnon spin-momentum locking.

\textit{Ferroelastically switchable anisotropic magnon spin transport}---The modification of the magnon bands across the ferroelastic transition profoundly alters magnon spin transport. The net spin conductivity, which originates from the unbalanced propagation of two chiral magnon modes (\cref{fig:CoTe2-1}(a)), is defined as $\sigma_{ij}^s=\sigma_{ij}^{\beta}-\sigma_{ij}^{\alpha}$ (see Sec.~S5 of SM \cite{supp}). This definition reflects the fact that the $\alpha$ and $\beta$ modes carry opposite spin angular momenta of $-\hbar$ and $+\hbar$, respectively. With increasing temperature, the contributions from both modes are enhanced due to the increased thermal magnon population, as shown in \cref{fig:CoTe2-1}(b-e). In the FA$_1$ phase, the net spin conductivity tensor is purely off-diagonal, with $\sigma^s_{xy}=1.92\times10^{-3}$~meV/K at 25~K (\cref{fig:CoTe2-1}(d)), substantially exceeding those ($\sim 0.3\times10^{-3}$~meV/K) reported for ferromagnetic \ch{CrPS4} and \ch{CrSBr} \cite{cuiAnisotropicMagnonTransport2025}. This off-diagonal transport stems from the counterflow of two chiral modes along the $y$-axis (\cref{fig:CoTe2-1}(a), top panel), driven by the distinct spatial anisotropies in the exchange interactions experienced by the two opposite-spin sublattices. Although both the $\alpha$ and $\beta$ magnons possess nonzero diagonal components ($\sigma_{ii}^\alpha = \sigma_{ii}^\beta \neq 0$ for $i=x, y$), their contributions to the net spin conductivities $\sigma^s_{ii}$ exactly cancel each other out, as exemplified by the $xx$ component in \cref{fig:CoTe2-1}(b). The ferroelastic transition to the FA$_2$ phase (\cref{fig:CoTe2-1}(a), bottom panel) reverses the sign of this $\beta$-$\alpha$ flow imbalance, resulting in $\sigma_{xy}^{s\prime}=-\sigma_{xy}^s$ (\cref{fig:CoTe2-1}(e)). Furthermore, while the diagonal components of both $\alpha$ and $\beta$ magnons change due to the interchange of principal axes, their exact cancellation persists in the FA$_2$ phase (\cref{fig:CoTe2-1}(d)).

Microscopically, the ferroelastically switchable spin transport fundamentally originates from the difference between the exchange interactions $J_{31}$ and $J_{32}$. To quantitatively understand this relationship, we calculate $\sigma^s_{xy}$ as a function of both temperature and the exchange difference $J_{31}-J_{32}$. As shown in \cref{fig:CoTe2-1}(f), $\sigma^s_{xy}$ exhibits a pronounced dependence on both parameters. At low temperatures, $\sigma^s_{xy}$ is suppressed due to the vanishing thermal population of magnons. When $J_{31}=J_{32}$, the system behaves as a conventional AFM devoid of chiral splitting, yielding zero spin conductivity. Conversely, increasing the magnitude of the exchange difference $|J_{31}-J_{32}|$ monotonically amplifies $|\sigma^s_{xy}|$, demonstrating that the AM character is the essential driver for magnon spin transport. Crucially, interchanging the values of $J_{31}$ and $J_{32}$ (i.e., reversing the sign of $J_{31}-J_{32}$) induces a sign reversal of $\sigma^s_{xy}$, consistent with the ferroelastic switching mechanism. These results underscore the critical role of AM chiral splitting in governing magnon spin transport and highlight its tunability via ferroelastic transitions.

The distinctive form of the spin conductivity tensor naturally gives rise to a pronounced anisotropy in magnon spin transport. Projecting the spin current onto the thermal-gradient direction $\hat{n}=(\cos\theta,\sin\theta)$, where $\theta$ is measured from the $x$-axis (see inset in \cref{fig:CoTe2-1}(g)), yields the longitudinal spin Seebeck and transverse spin Nernst conductivities, $\sigma^s_\parallel$ and $\sigma^s_\perp$. For \ch{CoTe2}, these components simplify to $\sigma^s_\parallel = \sigma^s_{xy}\sin 2\theta$ and $\sigma^s_\perp = \sigma^s_{xy}\cos 2\theta$, as detailed in Sec.~S6 of SM \cite{supp}. Consequently, both components oscillate with identical amplitudes as a function of $\theta$ (\cref{fig:CoTe2-1}(g)), reaching their respective maxima along the diagonal ($\theta=n\pi/2+\pi/4$, for integer $n$) and principal ($\theta=n\pi/2$) directions. Transitioning from the FA$_1$ to the FA$_2$ phase (\cref{fig:CoTe2-1}(h)) reverses both $\sigma^s_\parallel$ and $\sigma^s_\perp$, thereby fully inverting the spin current at any given angle $\theta$. In contrast, the anisotropic heat current (Fig.~S8 \cite{supp}) shows a sign reversal only in its transverse component, while the positive longitudinal component prevents a complete reversal of the total heat current. Our results thus demonstrate that spin transport in ferroelastic AMs can be deterministically controlled via ferroelastic switching. Crucially, such reversal of magnon spin currents can efficiently drive the magnetization switching of adjacent magnetic media via nonrelativistic spin splitting torques \cite{wangMagnetizationSwitchingMagnonmediated2019,hernandezEfficientElectricalSpin2021,karubeObservationSpinSplitterTorque2022a,baiObservationSpinSplitting2022a,huangManipulatingChiralSpin2024,zhangElectricalManipulationSpin2025,sarkarSpinsplitMagnonBands2025,guoMagneticMemoryDriven2025}, laying a foundation for magnonic logic gates and magnetic memories based on ferroelastic AMs.

\textit{Conclusion}---Based on symmetry analysis, we propose a general mechanism for the nonvolatile, field-free control of magnon spin current by harnessing intrinsic magnetoelastic coupling in 2D ferroelastic AMs. The ferroelastic transition deterministically reorients the crystal lattice and reshapes the magnetic exchange anisotropy. This process dynamically modulates nonrelativistic magnon spin transport, manifesting as a complete sign reversal of spin conductivities. Crucially, this mechanism requires neither external magnetic fields nor strong SOC, thereby simplifying device architectures and expanding the range of candidate materials. We validate this symmetry-driven mechanism through first-principles calculations on an AM \ch{CoTe2} monolayer. By establishing magnetoelastic coupling as a robust handle for controlling magnon spin transport, our findings can be broadly generalized to other ferroelastic AMs, facilitating the design of nonvolatile AFM magnonic devices.

\begin{acknowledgments}
	\textit{Acknowledgments}---We thank Dr. Yongqian Zhu for helpful discussion. This work is supported by National Key R\&D Program of China (2023YFA1406400).
\end{acknowledgments}

\end{document}